\documentclass[runningheads]{llncs}
\usepackage{xspace}
\usepackage{graphicx}
\usepackage{multirow}
\usepackage{amsmath}
\usepackage{amssymb}
\usepackage{fancyhdr}
\usepackage{hyperref}
\usepackage{marvosym}

\hypersetup{
    colorlinks=true, 
    linkcolor=blue,  
    filecolor=blue,  
    urlcolor=blue,   
    citecolor=blue,  
}
\newcommand{\our}{{WIA-LD2ND}\xspace}
\begin{document}
\title{WIA-LD2ND: Wavelet-based Image Alignment for Self-supervised Low-Dose CT Denoising}


\author{
Haoyu Zhao\inst{1} \and 
Yuliang Gu\inst{1} \and 
Zhou Zhao\inst{2} \and 
Bo Du\inst{1} \and \\ 
Yongchao Xu\inst{1}\textsuperscript{(\Letter)} \and 
Rui Yu\inst{3}
}

\institute{
    School of Computer Science, Wuhan University, Hubei, China \\ 
    \email{yongchao.xu@whu.edu.cn} \and
    School of Computer Science, Central China Normal University, \\ Hubei, China \and
    University of Louisville, Louisville, USA
}

% \institute{Anonymous Organization \\
% \email{**@******.***}}

\maketitle              % typeset the header of the contribution
\begin{abstract}
In clinical examinations and diagnoses, low-dose computed tomography (LDCT) is crucial for minimizing health risks compared with normal-dose computed tomography (NDCT). However, reducing the radiation dose compromises the signal-to-noise ratio, leading to degraded quality of CT images. To address this, we analyze LDCT denoising task based on experimental results from the frequency perspective, and then introduce a novel self-supervised CT image denoising method called \our, only using NDCT data. The proposed \our comprises two modules: Wavelet-based Image Alignment (WIA) and Frequency-Aware Multi-scale Loss (FAM). First, WIA is introduced to align NDCT with LDCT by mainly adding noise to the high-frequency components, which is the main difference between LDCT and NDCT. Second, to better capture high-frequency components and detailed information, Frequency-Aware Multi-scale Loss (FAM) is proposed by effectively utilizing multi-scale feature space. Extensive experiments on two public LDCT denoising datasets demonstrate that our \our, only uses NDCT, outperforms existing several state-of-the-art weakly-supervised and self-supervised methods. Source code is available at \url{https://github.com/zhaohaoyu376/WI-LD2ND}.

\end{abstract}

% ======================================================
% ======================================================
\section{Introduction}
Computed tomography (CT) has become a widely utilized tool in medical diagnosis. However, increased usage has raised concerns regarding potential risks associated with excessive radiation exposure~\cite{immonen2022use}. The widely recognized principle of ALARA (as low as reasonably achievable)~\cite{smith2009radiation} is extensively embraced to minimize exposure through strategies such as employing sparse sampling and reducing tube flux. Reducing the X-ray radiation dose, however, leads to poor-quality images with noticeable noise, which poses challenges for accurate diagnosis~\cite{diwakar2018review}. Therefore, the development of image denoising techniques~\cite{zhao2016loss} that can effectively handle CT modalities emerges as a critical and urgent need in clinical practice, to ensure both patient safety and diagnostic precision.

In recent years, advanced deep learning networks have proven to be highly effective in reducing noise in low-dose computed tomography (LDCT) than traditional denoising methods~\cite{dabov2007image,gu2014weighted}. Supervised denoising methods~\cite{lin2021artificial,li2022transformer}, such as CTformer~\cite{wang2023ctformer} and ASCON~\cite{chen2023ascon}, learn the end-to-end mapping from low-dose to normal-dose CT images. Generative adversarial networks (GANs) are also utilized in LDCT denoising task, which do not need paired data, but lots of unpaired data for training.\cite{park2020contrastive,bera2023axial,kwon2021cycle,jing2023inter}.

\begin{figure}[t]
  \centering
  % 左边的大图 (a)
  \begin{minipage}{0.5\linewidth}
    \centering
    \includegraphics[width=1.0\linewidth]{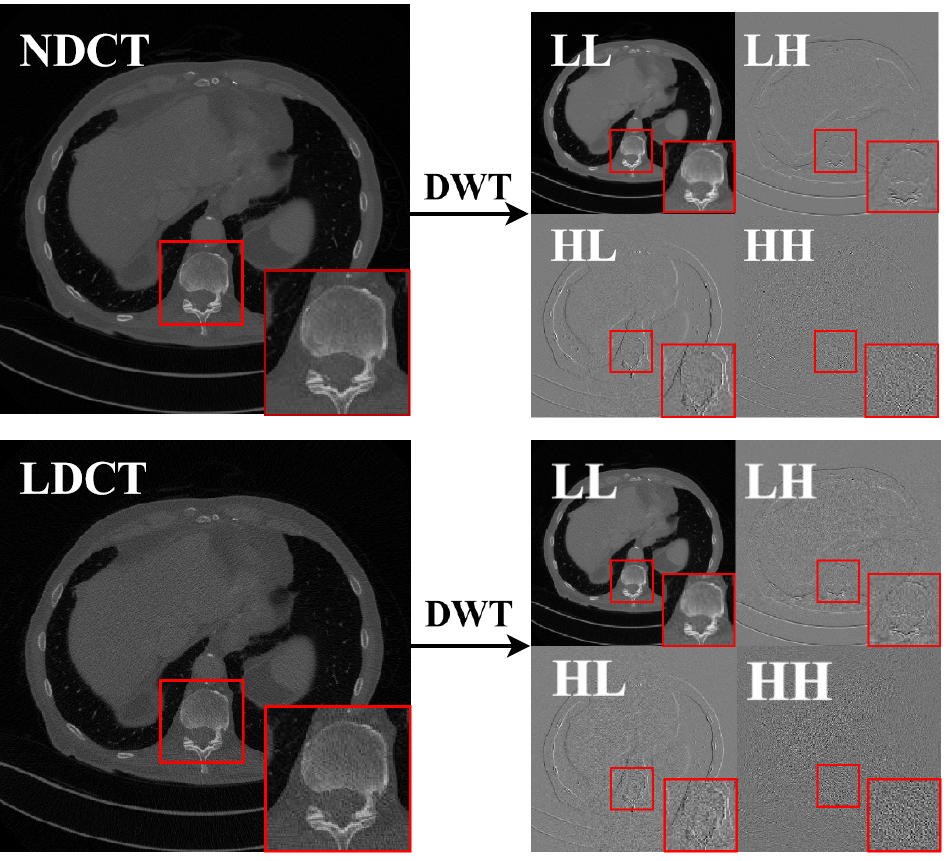}\\
    (a) Visualization of the results of NDCT and LDCT after Wavelet transform results
  \end{minipage}%
  \hfill
  % 右边的四个小图 (b, c, d, e)
  \begin{minipage}{0.49\linewidth}
    \centering
    % 上面两个小图 (b, c)
    \begin{minipage}{0.49\linewidth}
      \centering
      \includegraphics[width=1.0\linewidth]{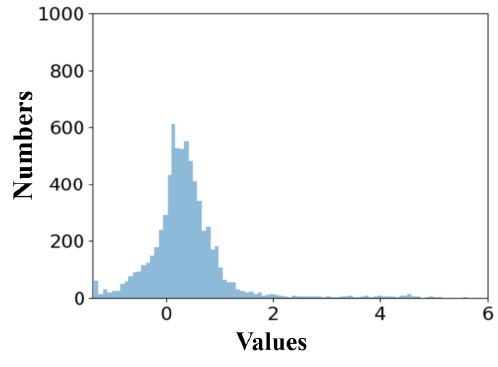}\\
      (b) LF components of NDCT
    \end{minipage}%
    \begin{minipage}{0.49\linewidth}
      \centering
      \includegraphics[width=1.0\linewidth]{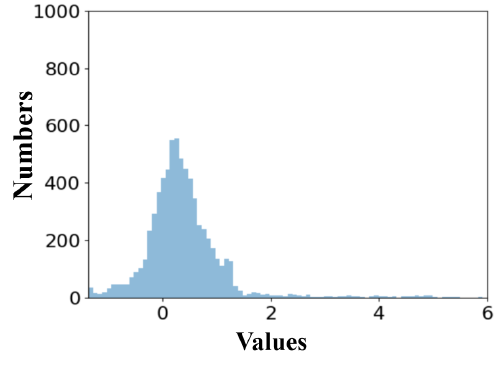}\\
      (c) LF components of LDCT
    \end{minipage}
    % 下面两个小图 (d, e)
    \begin{minipage}{0.49\linewidth}
      \centering
      \includegraphics[width=1.0\linewidth]{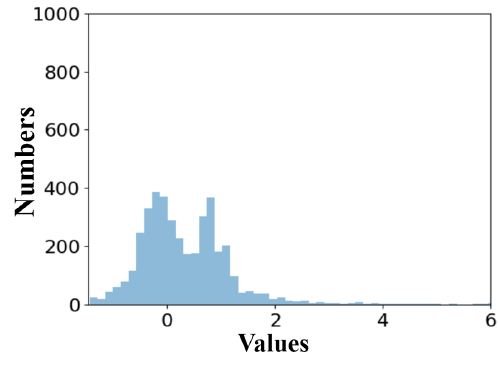}\\
      (d) HF components of NDCT
    \end{minipage}%
    \begin{minipage}{0.49\linewidth}
      \centering
      \includegraphics[width=1.0\linewidth]{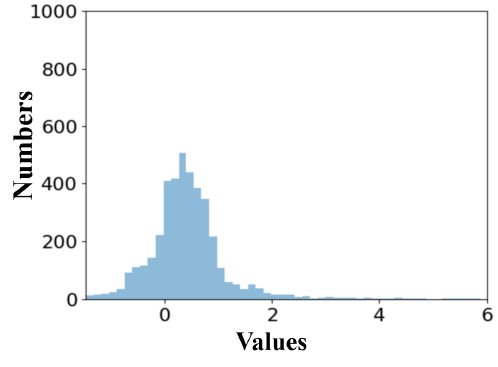}\\
      (e) HF components of LDCT
    \end{minipage}
  \end{minipage}
  \caption{(a) Visualization of results of NDCT and LDCT after Discrete Wavelet Transform (DWT). The primary differences between NDCT and LDCT are at the high frequency components $[LH, HL, HH]$. (b-c) Visualize the normalized low-frequency (LF) component $LL$ features of NDCT and LDCT, while (d-e) display the normalized high-frequency (HF) component $[LH, HL, HH]$ features. We adopt the first residual block of pre-trained ResNet-18~\cite{he2016deep} to extract image features.}
  \label{fig:frequency_difference}
\end{figure}

Despite impressive results, these methods encounter challenges as they require both LDCT and NDCT images~\cite{liu2022learning}, either paired images or a large amount of unpaired images, which are often unavailable in practice due to high costs, privacy, and ethical concerns. Therefore, it is essential to develop self-supervised methods that harness the potent capabilities of deep neural networks while minimizing the need for extensive labeled data. Several self-supervised methods have been proposed for LDCT denoising, including but not limited to Blin2Unblind~\cite{wang2022blind2unblind}, Noise2Sim~\cite{niu2022noise}, Neighbor2Neighbor~\cite{huang2022neighbor2neighbor} and FIRE~\cite{long2023full} among others~\cite{ulyanov2018deep,jing2022training,wang2022blind2unblind}. However, these methods primarily concentrate on spatial domain information, overlooking the critical importance of frequency domain details. 
% As shown in Fig~\ref{fig:frequency_difference}, the main difference between LDCT and NDCT lays in high frequency components. 
The crucial distinction between low-dose CT and normal-dose CT in high-frequency components (see Fig.~\ref{fig:frequency_difference}) is not well explored.

In this paper, we design a novel self-supervised LDCT denoising method, only using NDCT data, called \our. We first analyze LDCT denoising task from the frequency perspective and then propose a module called Wavelet-based Image Alignment (WIA), which aligns LDCT with NDCT by mainly adding noise to the high-frequency components of both LDCT and NDCT. We also propose a module called Frequency-Aware Multi-scale Loss (FAM) to capture high-frequency components in multi-scale feature space.

Our \our offers three major contributions as follows: 
1) We analyze the LDCT denoising task from a frequency perspective, offering novel insight into its optimization. 2) We introduce a simple and efficient module to align NDCT and LDCT, facilitating self-supervised learning. 3) We propose a frequency-aware multi-scale loss, enabling the reconstruction network to effectively handle high-frequency components. 

% ===================================================
% ===================================================

\begin{figure}[t]
  \centering
  \includegraphics[width=0.95\linewidth]{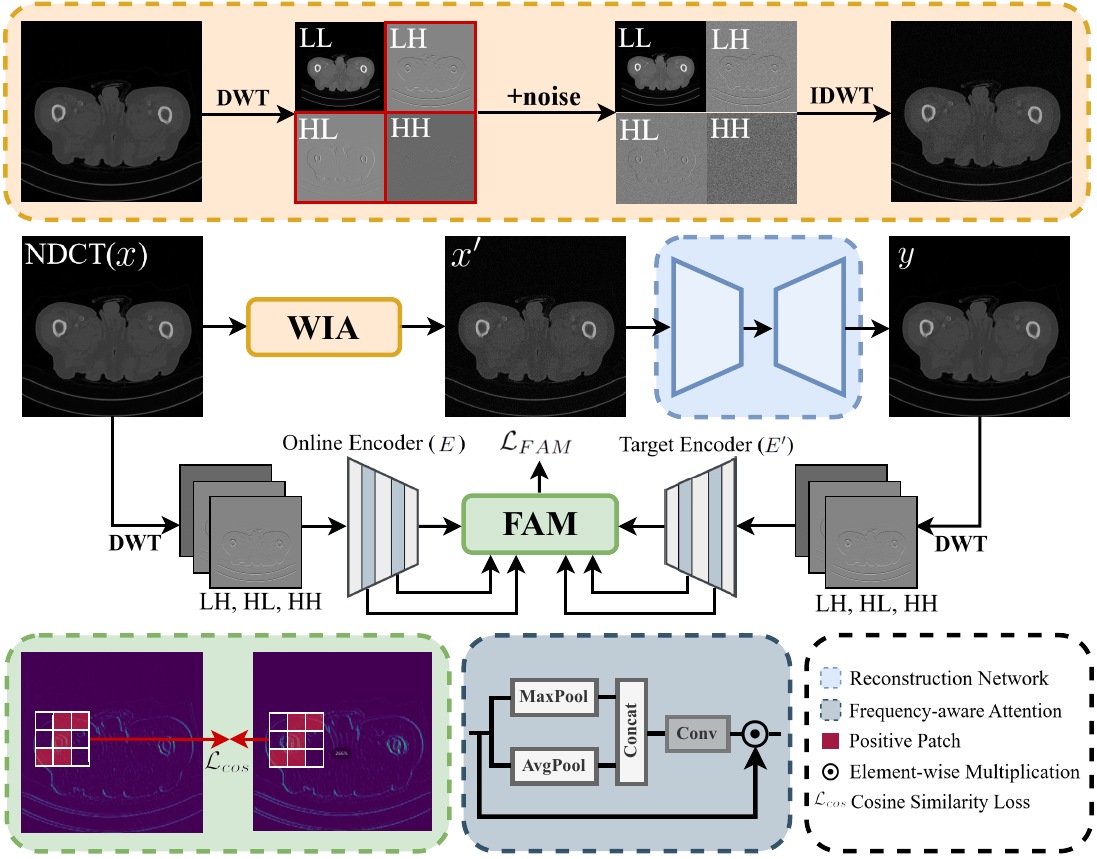}
   \caption{Overview of our proposed \our. NDCT image $x$ is passed through the WIA and then fed into the reconstruction network. The high-frequency components of the denoised CT $y$ and the input image $x$ are both fed into FAM to compute the loss capturing high-frequency components in multi-scale feature space.}
   \label{fig:pipline}
\end{figure}

% ===================================================
% ===================================================

\section{Method}
% \noindent
% \textbf{Model Overview.}
Figure~\ref{fig:pipline} presents the overview of the proposed \our, comprising of two novel modules: Wavelet-based Image Alignment (WIA) and Frequency-Aware Multi-scale Loss (FAM). NDCT image $x$ is passed through the WIA to destroy some high-frequency components and then fed into the reconstruction network. 
The high-frequency components of the reconstructed CT $y$ and input $x$, [$y_{LH}$, $y_{HL}$, $y_{HH}$] and [$x_{LH}$, $x_{HL}$, $x_{HH}$], are relatively flat and hard to capture~\cite{gou2023rethinking}, as shown in Fig.~\ref{fig:frequency_difference} (a). 
To address this challenge, we propose integrating these high-frequency components into encoders to compute a loss $\mathcal{L}_{FAM}$ within the feature space to enhance the capability of the reconstruction network in capturing high-frequency components and detailed information more effectively.
During training, we employ an alternating learning strategy to optimize the reconstruction network and FAM to improve learning efficiency and accuracy of results, which is similar to GAN-based methods~\cite{zhu2017unpaired}. 
We begin by analyzing the LDCT denoising task from a novel perspective, followed by detailed introduction of the two proposed modules.

% ===================================================
% ===================================================

% \textbf{Analysis of CT Denoising From Frequency Perspective.}
\subsection{Analysis of LDCT Denoising From Frequency Perspective}
\label{sec:analysis}
Images contain different frequency ranges and spatial locations information. The Discrete Wavelet Transform (DWT), using the Haar wavelet as in~\cite{liu2020wavelet}, is selected for frequency analysis for its simplicity and efficiency. DWT is commonly employed in the field of computer vision and offers a straightforward and computationally effective technique for dividing the input image into low-frequency sub-band and high-frequency sub-bands. It has four filters, ${LL^T}$, ${LH^T}$, ${HL^T}$ and ${HH^T}$, demonstrating the texture, horizontal details, vertical details, and diagonal information respectively~\cite{yao2022wave}, in which low and high pass filters are: 

\begin{equation}
L^T = \frac{1}{\sqrt{2}}[1,1], 
H^T = \frac{1}{\sqrt{2}}[-1,1]
\label{wavelet1}
\end{equation}

As shown in Fig.~\ref{fig:frequency_difference} (a), after DWT, the main differences between normal-dose CT (NDCT) and low-dose CT (LDCT) are observed in the high-frequency sub-images $[LH, HL, HH]$, with little difference in the $LL$. Fig.~\ref{fig:frequency_difference} (b-e) further support our conclusion, demonstrating that LDCT and NDCT have significant differences in the high-frequency components at the feature space, while differences in the low-frequency components are comparatively minor. In Fig.~\ref{fig:wia} (a-c) and Fig.~\ref{fig:result}, we find that previous LDCT denoising methods such as BM3D~\cite{dabov2007image} performs poorly at reconstructing high-frequency components.

Based on these observations, we conclude that high-frequency components should be the main focus for LDCT denoising, where previous methods falter the most. This highlights the critical necessity for improved techniques in handling high-frequency components during the denoising process.

\subsection{Wavelet-based Image Alignment}

\begin{figure}[t]
  \centering
  \begin{minipage}{0.175\linewidth}
    \centering
    \includegraphics[width=\linewidth]{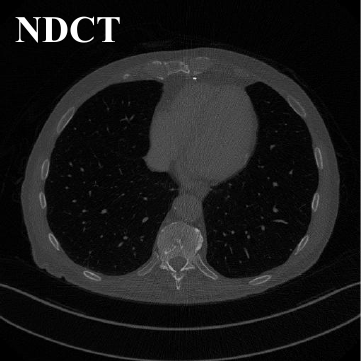}\\
    (a)
  \end{minipage}
  \begin{minipage}{0.175\linewidth}
    \centering
    \includegraphics[width=\linewidth]{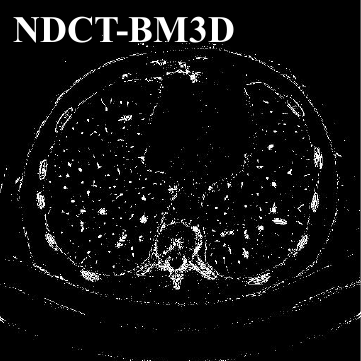}\\
    (b)
  \end{minipage}
  \begin{minipage}{0.175\linewidth}
    \centering
    \includegraphics[width=\linewidth]{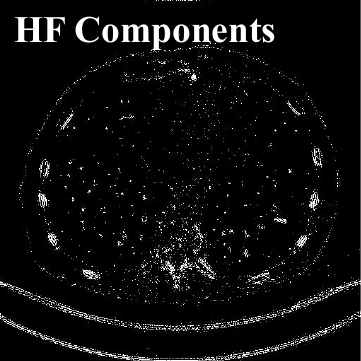}\\
    (c)
  \end{minipage}
  \begin{minipage}{0.221\linewidth}
    \centering
    \includegraphics[width=\linewidth]{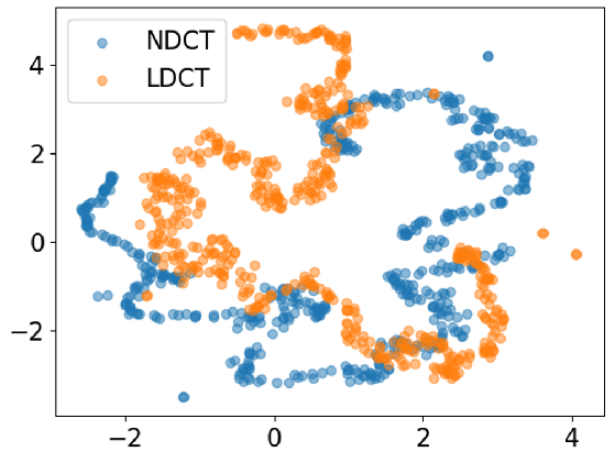}\\
    (d)
  \end{minipage}
  \begin{minipage}{0.221\linewidth}
    \centering
    \includegraphics[width=\linewidth]{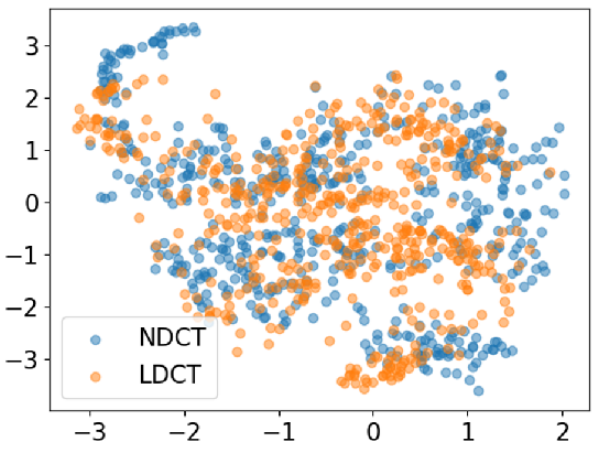}\\
    (e)
  \end{minipage}
  \caption{(a-c) Visualization of NDCT, residual between NDCT and result of BM3D~\cite{dabov2007image} (a classical denoising method), and high-frequency components in spatial domain. The residual is converted into a clean binary image for clarity. We filter high-frequency band from image and then convert the result into a binary image. (d-e) Visualization of the tSNE images of feature distribution on the NDCT, LDCT, and their respective transformations after applying WIA on Mayo-2016 dataset. We adopt the first residual block of pre-trained ResNet-18 to extract image features.}
  \label{fig:wia}
\end{figure}

According to previous analysis, we employ the Discrete Wavelet Transform (DWT) to decompose the input image $x$ into two sets of components: low-frequency component denoted as $x_{LL}$, which captures and preserves the smooth surface and texture information, and high-frequency components denoted by [$x_{LH}$, $x_{HL}$, $x_{HH}$]. These high-frequency components are essential for capturing intricate texture details, representing the primary distinctions between low-dose CT (LDCT) and normal-dose CT (NDCT) images. Therefore, we make them similar in a simple and effective way, by mainly adding Gaussian noise into the high-frequency components of NDCT and LDCT images.

\begin{equation}
\begin{aligned}
x_{LL}' = x_{LL} + noise_{LL}, \quad x_{LH}' = x_{LH} + noise_{LH}, \\
x_{HL}' = x_{HL} + noise_{HL}, \quad x_{HH}' = x_{HH} + noise_{HH},
\label{wavelet2}
\end{aligned}
\end{equation}

where $noise_{LL}, noise_{LH}, noise_{HL}, noise_{HH}$ follow Gaussian distributions with mean 0 and variance $\sigma_{LL}^2, \sigma_{LH}^2, \sigma_{HL}^2, \sigma_{HH}^2$, respectively. Notably, $\sigma_{LH}, \sigma_{HL}$, and $\sigma_{HH}$ are larger than $\sigma_{LL}$. We then conduct inverse DWT (IDWT) on the modified components [$x_{LL}'$,$x_{LH}'$, $x_{HL}'$, $x_{HH}'$] to reconstruct $x'$. As shown in Fig.~\ref{fig:wia} (d-e), after applying WIA module, NDCT and LDCT images share the same feature space, indicating successful alignment.

WIA eliminates the need for paired data. 
% Instead, we only require the NDCT images and let the model learn the transition from NDCT after WIA $x'$ to NDCT $x$ to denoise, thereby facilitating self-supervised learning.
Instead, we only require NDCT images for training a model to denoise from $x'$ (NDCT after WIA) to $x$ (original NDCT), thereby facilitating self-supervised learning.

% ===================================================
% ===================================================

% ===================================================
% ===================================================

\subsection{Frequency-Aware Multi-scale Loss}
High-frequency components play a crucial role in low-dose CT images denoising, as analyzed in Sec.~\ref{sec:analysis}. However, the CNN-based and Transformer-based models tend to focus primarily on low-frequency representations, making it difficult for models to capture the high-frequency components~\cite{gou2023rethinking}. Therefore, we design Frequency-Aware Multi-scale Loss (FAM) which is to focus the network more on the high-frequency components and detail information of the images. 

% ===================================================
% ===================================================

The high-frequency components of denoised CT $y$ and NDCT $x$, [$y_{LH}$, $y_{HL}$, $y_{HH}$] and [$x_{LH}$, $x_{HL}$, $x_{HH}$], are fed into Online Encoder $E$ and Target Encoder $E'$, respectively, both using the same lightweight architecture. Following previous studies~\cite{grill2020bootstrap,he2020momentum}, the parameters of the Target Encoder $E'$ are an exponential moving average of the parameters in the Online Encoder $E$. The process is as follows:

\begin{equation}
\begin{aligned}
f_1, f_2, f_3 = E(x_{LH}, x_{HL}, x_{HH}), f_1', f_2', f_3' = E'(y_{LH}, y_{HL}, y_{HH}).
\end{aligned}
\label{encoder}
\end{equation}

We introduce a Frequency-aware Attention mechanism in the encoders, designed to selectively emphasize or de-emphasize areas within the input feature map based on their frequency content, as illustrated in Fig.~\ref{fig:pipline}
% Given the feature map $\{f_n\}_{n \in \{1,2,3\}}$, we perform max and average pooling to capture prominent features. 
Given the input feature map ${f_n}_{\{n=1,2,3\}}$, we apply max and average pooling to extract prominent features. We then concatenate them and pass through a convolutional layer with Sigmoid activation to generate spatial attention weights.

Unlike previous studies~\cite{chen2023ascon,yun2022patch}, our approach segments multi-scale features extracted from specific layers into patches $f_n^{(i)}$. We then select patches that are most similar to their adjacent counterparts, focusing on those that exhibit shared structural characteristics closely associated with high-frequency components. To this end, we use the cosine similarity $s$ on the feature space:

\begin{equation}
\begin{aligned}
\text{s}(i,j) = f^{(i)^\top} f^{(j)} / \| f^{(i)}\|_{2} \| f^{(j)}\|_{2}.
\end{aligned}
\label{similarity}
\end{equation}

We then select the similar feature patches $\{ f_n^{(j)} \}_{j \in P^{(i)}}$, where $P^{(i)}$ is a set of feature patch indices of top-4 patches. For the $f'^{(i)}$ from the Online Network, we select the same positive feature patches $P^{(i)}$. We then aggregate positive patches $\{ f_n^{(j)} \}_{j \in P^{(i)}}$ and $\{ f_n'^{(j)} \}_{j \in P^{(i)}}$ using global average pooling (GAP) and multi-layer perceptron (MLP) yielding $g$ and $g'$, respectively. Finally, the Frequency-Aware Multi-scale Loss $\mathcal{L}_{FAM}$ is given by:
%which is important for self-learning.

\begin{equation}
\begin{aligned}
\mathcal{L}_{FAM} = \|g-g'\|_{2}^{2}.
\end{aligned}
\label{high_level}
\end{equation}

The final loss is defined as $\mathcal{L} = \mathcal{L}_{pixel}(x, y) + \lambda \mathcal{L}_{FAM}$, where $\mathcal{L}_{pixel}$ consists of two common supervised losses: MSE and SSIM, defined as $\mathcal{L}_{pixel}(x, y) = \mathcal{L}_{MSE}(x, y) +\mathcal{L}_{SSIM}(x, y)$. $\lambda$ is set to 0.01 in this paper.

% ===================================================
% ===================================================

\begin{table}[t]
\centering
\footnotesize
\caption{Performance comparison on the Mayo-2016~\cite{mccollough2017low} and Mayo-2020~\cite{moen2021low} datasets. The best result is in \textbf{bold}, and the second best is \underline{underlined}.}
\scriptsize
\setlength{\tabcolsep}{6pt}
\begin{tabular}{lcccccc}
\hline
\multirow{2}{*}{\textbf{Methods}} & \multicolumn{2}{c}{\textbf{Mayo-2016}} & \multicolumn{2}{c}{\textbf{Mayo-2020}} & \multicolumn{2}{c}{\textbf{Avg}}\\
\cline{2-7} & PSNR & SSIM & PSNR & SSIM & PSNR & SSIM\\
 \hline
 % LDCT        & 32.25 & 72.16 &  &  \\
 % \hline
 BM3D \tiny{(TIP2007)}~\cite{dabov2007image}               & 35.50 & 86.95 & 37.67 & 89.53 & 36.59 & 88.24 \\
 DIP \tiny{(CVPR2018)}~\cite{ulyanov2018deep}               & 37.25 & 85.94 & 40.16 & 95.89 & 38.71 & 90.92 \\
 % RED-CNN     &  &  &  &   \\
 Noise2Sim \tiny{(TMI2022)}~\cite{niu2022noise}             & \textbf{38.51} & \textbf{90.15} & 39.16 & 90.90  & 38.84 & 90.53 \\
 Blind2Unblind \tiny{(CVPR2022)}~\cite{wang2022blind2unblind}& 35.84 & 81.15 & 40.96 & 94.84  & 38.40 & 88.00 \\
 Neighbor2Neighbor \tiny{(TIP2022)}~\cite{huang2022neighbor2neighbor} & 35.27 & 86.79 & 36.78 & 94.06 & 36.03 & 90.43 \\
 ZS-N2N \tiny{(CVPR2023)}~\cite{mansour2023zero}  & 38.10 & 87.64 & \underline{44.00} & 97.06 & \underline{41.05}  & 92.35 \\
 CycleGAN \tiny{(ICCV2017)}~\cite{zhu2017unpaired}           & 37.68 & 89.22 & 40.18 & \underline{97.93}  & 38.93 & 93.58 \\
 CUT \tiny{(ECCV2020)}~\cite{park2020contrastive}            & 37.96 & 89.93 & 41.11 & 97.61  & 39.54 & \underline{93.77} \\
 \hline
 \our (ours)    & \underline{38.15} & \underline{90.00} & \textbf{44.64} & \textbf{98.31} & \textbf{41.40} & \textbf{94.16} \\
\hline
\end{tabular}
\label{tab:experiment}
\end{table}

% ===================================================
% ===================================================

\begin{figure}[t]
  \centering
  \includegraphics[width=1\linewidth]{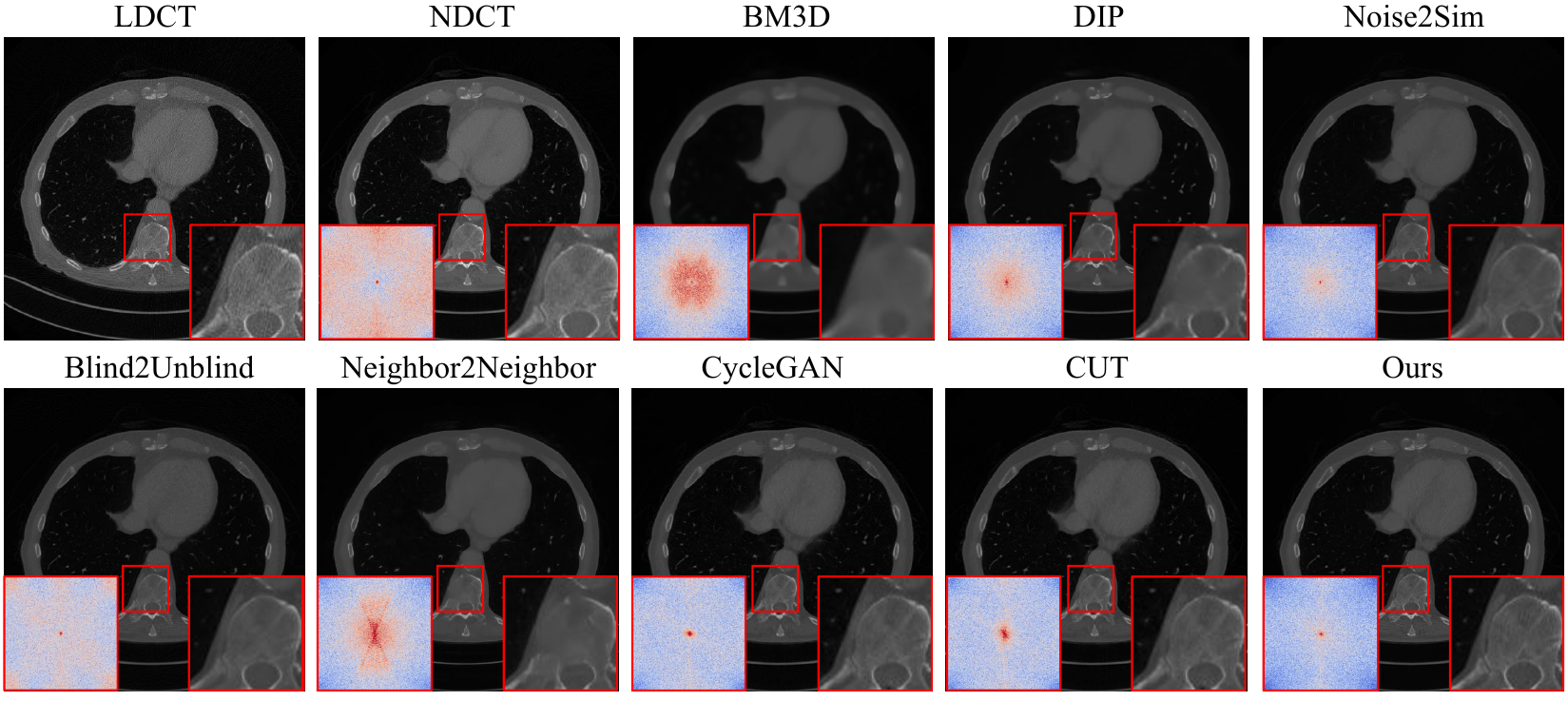} % 调整图像宽度
   \caption{Qualitative comparison of different methods on the Mayo-2020 dataset~\cite{moen2021low}.}
   \label{fig:result}
\end{figure}

% ===================================================
% ===================================================

\section{Experiments}
\subsection{Dataset and Training Details}

We conduct experiments on two public LDCT denoising datasets, Mayo-2016\footnote{\url{https://ctcicblog.mayo.edu/2016-low-dose-ct-grand-challenge/}} and Mayo-2020\footnote{\url{https://wiki.cancerimagingarchive.net/pages/viewpage.action?pageId=52758026}}, from the NIH AAPM-Mayo Clinic Low-Dose CT Grand Challenge~\cite{mccollough2017low,moen2021low}. We select 5410 image pairs (512×512) from 9 patients in Mayo-2016 for training and 526 for testing. We select the reconstruction parameter combination of \{1mm, D45\}. From Mayo-2020, 2082 pairs (512×512) from 12 patients are used for training, with 672 pairs from 4 patients for testing.

In all our experiments, we only use NDCT and choose a backbone identical to that used in~\cite{zhu2017unpaired,park2020contrastive}, along with employing the same data augmentation strategy as in~\cite{chen2023ascon}. For Mayo-2016, we set $\sigma_{LL}=100$, $\sigma_{LH}=200$, $\sigma_{HL}=200$, and $\sigma_{HH}=150$ in Eq.~\eqref{wavelet2}. For Mayo-2020, the noise variances are $\sigma_{LL}=25$, $\sigma_{LH}=50$, $\sigma_{HL}=50$, and $\sigma_{HH}=50$. We employ the Adam optimizer with the momentum parameters as $\beta_1$ = 0.9, $\beta_2$ = 0.99 and initial learning rate $1.0 \times 10^{-4}$. Our network is trained over 200 epochs using a single NVIDIA GeForce RTX 3090.

% ===================================================
% ===================================================

\subsection{Experiments Results}
To evaluate the denoising efficacy of our \our, we conduct comparative experiments against various denoising methods. The comparisons include traditional methods like BM3D~\cite{dabov2007image}, self-supervised methods including DIP~\cite{ulyanov2018deep}, Noise2Sim~\cite{niu2022noise}, Blind2Unblind~\cite{wang2022blind2unblind}, Neighbor2Neighbor~\cite{huang2022neighbor2neighbor} and ZS-N2N~\cite{mansour2023zero} as well as weakly-supervised methods such as CycleGAN~\cite{zhu2017unpaired} and CUT~\cite{park2020contrastive}. We use two widely-adopted metrics, namely peak signal-to-noise ratio (PSNR) and structural similarity index measure (SSIM) to evaluate the performance.

Table~\ref{tab:experiment} shows that our \our, only using NDCT images, can achieve good performance on both the Mayo-2016 and Mayo-2020 datasets. 
Compared to the latest state-of-the-art self-supervised and weakly-supervised methods, our \our achieves significant performance improvements.

Figure~\ref{fig:result} presents the reconstruction results, with the subplots in the bottom left corner of the pictures showing the noise power spectrum (NPS), where blue indicates it is closer to the normal-dose CT. As illustrated, our \our achieves the best results, reconstructing the most detailed information and exhibiting the bluest NPS. Conversely, the results of BM3D and DIP are over-smoothed and compromised with structured artifacts. Additionally, other deep-learning-based methods tend to remove noise aggressively. Our \our prioritizes the preservation of informative details.

% ===================================================
% ===================================================

\begin{table}[t]
\centering
\footnotesize
\caption{Ablation studies are conducted to validate the effectiveness of each module on the Mayo-2016~\cite{mccollough2017low} and Mayo-2020~\cite{moen2021low} datasets. WIA\textsuperscript{*} represents directly adding Gaussian noise to NDCT, while FAM\textsuperscript{*} denotes the computation of high-frequency components of $y$ and $x$ at the feature level by directly employing the MSE loss.}
\setlength{\tabcolsep}{5pt}
\scriptsize
\begin{tabular}{lcccccccc} 
\hline 
\multirow{2}{*}{\textbf{Methods}} & \multirow{2}{*}{\textbf{\#Params}} & \multicolumn{2}{c}{\textbf{Mayo-2016}} & \multicolumn{2}{c}{\textbf{Mayo-2020}} & \multicolumn{2}{c}{\textbf{Avg}} \\
\cline{3-8} 
 &  & PSNR & SSIM & PSNR & SSIM & PSNR & SSIM\\ 
\hline
Baseline  & 11.37M & 34.26 & 78.03 & 40.32 & 96.99 & 37.29 & 87.51\\
Baseline + WIA\textsuperscript{*} & 11.37M  & 35.49 & 87.73 & 42.00 & 98.26 & 38.75 & 93.00\\
Baseline + WIA & 11.37M  & 37.85 & 89.77 & 42.73 & 98.26 & 40.29 & 94.02\\
\hline
Baseline + FAM\textsuperscript{*} & 13.83M & 34.00 & 79.12 & 41.97 & 98.02 & 37.99 & 88.69 \\
Baseline + FAM & 13.83M  & 34.78 & 80.27 & 42.46 & 98.25 & 38.62 & 89.15\\
\hline
\our & 13.83M  & \textbf{38.15} & \textbf{90.00} & \textbf{44.64} & \textbf{98.31} & \textbf{41.40} & \textbf{94.16} \\
\hline
\end{tabular}
\label{tab:ablation}
\end{table}

% ===================================================
% ===================================================

\medskip
\noindent
\textbf{Ablation Studies}.
% \subsection{Ablation Studies}
To evaluate the effectiveness of our proposed modules, including WIA and FAM, we conduct ablation experiments on Mayo-2016~\cite{mccollough2017low} and Mayo-2020~\cite{moen2021low}. The results are shown in Table~\ref{tab:ablation}. WIA\textsuperscript{*} represents adding noise directly to NDCT, while FAM\textsuperscript{*} involves the computation of MSE loss directly for the feature of high-frequency components. Our designs achieve better performance than their variants. It reveals that both WIA and FAM are well-designed and contribute to performance gains. More ablation studies on additional hyperparameters and noise parameters are available in the supplementary materials. \our incurs an increase of 2.46M parameters compared to the baseline. Considering the significant performance improvement over the baseline model \textbf{without any extra inference time}, this slight increase in training cost is acceptable.

% ===================================================
% ===================================================

\section{Discussion and Conclusion}
In this paper, we analyze the LDCT denoising task from a novel perspective and propose a self-supervised method called \our. This method only utilizes NDCT images and incorporates two novel modules: Wavelet-based Image Alignment (WIA), which aligns NDCT and LDCT by destroying some high-frequency components, and Frequency-Aware Multi-scale Loss (FAM), which enhances the reconstruction network's ability to capture high-frequency components and detailed information, thus improving denoising performance. Extensive experimental results demonstrate the superior performance and the effectiveness of our designs. It is noteworthy that \our increases the number of parameters by 2.46M compared to the baseline during training, without requiring extra inference time. Exploring LDCT denoising from a frequency perspective presents a promising direction for future research.

\bibliographystyle{splncs04}
\bibliography{Paper-0783}

\end{document}